\documentstyle[aps,epsf]{revtex}

\global\firstfigfalse  
\global\firsttabfalse  

\newcommand{\drawsquare}[2]{\hbox{%
\rule{#2pt}{#1pt}\hskip-#2pt
\rule{#1pt}{#2pt}\hskip-#1pt
\rule[#1pt]{#1pt}{#2pt}}\rule[#1pt]{#2pt}{#2pt}\hskip-#2pt
\rule{#2pt}{#1pt}}

\newcommand{\fund}{\drawsquare{6.5}{0.4}}
\newcommand{\afund}{\overline{\fund}}

\newcommand{\beq}{\begin{eqnarray}}
\newcommand{\eeq}{\end{eqnarray}}

\newcommand{\Kahler}{K\"ah\-ler }
\newcommand{\scr}[1]{{\cal{#1}}}
\renewcommand{\d}{\partial}
\newcommand{\mybar}[1]%
	{{\kern 0.8pt\overline{\kern -0.8pt#1\kern -0.8pt}\kern 0.8pt}}

\newcommand{\roughly}[1]%
	{{\mathrel{\raise.3ex\hbox{$#1$\kern-.75em\lower1ex\hbox{$\sim$}}}}}
\newcommand{\myint}{\int\mkern-5mu}
\newcommand{\lsim}{\mathrel{\roughly<}}
\newcommand{\gsim}{\mathrel{\roughly>}}
\newcommand{\avg}[1]{\langle #1 \rangle}
\newcommand{\Avg}[1]{\left\langle #1 \right\rangle}
\newcommand{\sfrac}[2]{{\textstyle\frac{#1}{#2}}}
\newcommand{\eql}[1]{\label{eq:#1}}
\newcommand{\eq}[1]{(\ref{eq:#1})}  
\newcommand{\Eq}[1]{Eq.~\eq{#1}}  	 

\renewcommand{\Re}{\mathop{\rm Re}}
\renewcommand{\Im}{\mathop{\rm Im}}

\newcommand{\Journal}[4]{{#1}\ {\bf #2}, #3 (#4)}

\newcommand{\NPB}[3]{\Journal{Nucl.\ Phys.}{B#1}{#2}{#3}}

\newcommand{\PLB}[3]{\Journal{Phys.\ Lett.}{#1B}{#2}{#3}}

\newcommand{\PRD}[3]{\Journal{Phys.\ Rev.}{D#1}{#2}{#3}}

\newcommand{\PTP}[3]{\Journal{Prog.\ Theor.\ Phys.}{#1}{#2}{#3}}

\begin{document}

\baselineskip 14pt

\title{Single Sector Supersymmetry Breaking}
\author{Markus A. Luty\footnote{Sloan Fellow}}
\address{Department of Physics,
University of Maryland\\
College Park, Maryland 20742, USA\\
{\tt mluty@physics.umd.edu}}
\author{John Terning\thanks{Talk presented by J. Terning at DPF '99, Los
Angeles.}}   \address{Department of Physics,
University of California\\
and\\
Theory Group, Lawrence Berkeley Laboratory\\
Berkeley, California 94720, USA\\
{\tt terning@alvin.lbl.gov}}
\maketitle
\vspace{-202pt}
\begin{center}
  \hfill  UCB-PTH-99/09 \\
~{} \hfill    LBNL-42988 \\
\end{center}
\vspace{158pt}

\begin{abstract}
We review recent work on
realistic models that break supersymmetry dynamically and give
rise to composite quarks and leptons, all in a single
sector.  These models have a completely natural suppression of flavor-changing
neutral currents,  and the hierarchy of
Yukawa couplings is explained by the dimensionality of composite states.
The generic signatures are unification of scalar
masses with different quantum numbers at the compositeness scale, and
lighter gaugino, Higgsino, and third-generation sfermion masses.
\end{abstract}

\section{Introduction}
One of the most exciting results of the recent progress in
understanding strongly coupled
supersymmetric gauge theories \cite{Seiberg} is that it
allows the exploration of new possibilities for the realization of
supersymmetry (SUSY) in nature.
The most important example is the use of dynamical SUSY
breaking to explain the origin of the SUSY breaking scale.
In recent years, large classes of SUSY
gauge theories have been discovered
that exhibit dynamical SUSY breaking through a
variety of different mechanisms \cite{break,LutyTerning},
and realistic models have been built using these theories
as building blocks in both (super)gravity-mediated
and gauge-mediated frameworks.

In these conventional approaches to SUSY model building, SUSY breaking
arises in a separate sector consisting of fields that are neutral under the
standard model (SM) gauge group, and SUSY breaking is communicated to the
observable fields by messenger (gauge or gravitational) interactions.
It is clearly important to know whether such a `modular' structure is required
in order for SUSY to be the solution of the hierarchy problem,
or if simpler models without disjoint sectors are possible.
In \cite{ALT,SSSB}, realistic models were constructed that do not
require a separate SUSY breaking sector.
In these models, SUSY is broken dynamically by strongly coupled
fields that are charged
under the SM gauge group, giving rise to composites
with the quantum numbers of quarks and leptons.
The composite sfermions have
SUSY breaking masses induced directly by the strong dynamics, while
the masses of the composite fermions are protected by
unbroken chiral symmetries.
The masses of the elementary gauginos and sfermions arise
from gauge mediation (via the composite scalars);
they are therefore smaller than the
composite scalar masses, which are necessarily in the range of
1--10 TeV.

If we make the simplest assumption that the first two generations are
composite while the third is elementary, we automatically gain a partial
understanding of the observed hierarchy of fermion masses.
The reason is that all Yukawa couplings involving composite
states must arise from higher-dimension operators in the fundamental theory,
and are thus suppressed, while the top Yukawa coupling can
be order one.
A highly non-trivial feature of this scenario is that it does not lead
to excessive flavor-changing neutral currents (FCNC's) from squark
non-degeneracy even if the flavor sector has no flavor symmetry.
This is because the strong composite dynamics is flavor-blind,
and so the composite scalar masses are
degenerate to high accuracy,
with small corrections due to perturbative flavor-breaking couplings.

Such scenarios for single sector SUSY breaking have several
interesting generic phenomenological implications. First, as mentioned
above, gaugino and stop masses will be much smaller
than the masses of composite squarks and sleptons.
Second, the composite scalar masses unify at the compositeness
scale.
Third, if we assume that the Yukawa interactions are generated by new
physics at a flavor scale above the compositeness scale without special
flavor symmetries, predictions for flavor-changing processes such as
$\mu \to e\gamma$ are plausibly within experimental reach.

 We have constructed large classes of supersymmetric gauge
theories \cite{SSSB} with the non-perturbative dynamics required for this
kind of model-building.
At low energies the models either
confine (like the models of \cite{ALT}),
have conformal fixed points, or are magnetically free.
All of the models discussed here have only local SUSY breaking minima.
Also, they are not calculable, and  require
dynamical assumptions to be phenomenologically viable.
However, in many of the models the existence of a local SUSY
breaking minimum with composite fermions
can be established without dynamical assumptions.
This shows that the combination of compositeness and SUSY breaking is
not exotic,
and suggests that further exploration of the connection
between these phenomena is worthwhile.

\section{Mass Scales and Phenomenology}
In this Section, we describe the most important qualitative features
of single sector models.
We will then  focus on three example models:
a `meson' model where the first two generations correspond to dimension 2
operators; a `dimensional hierarchy' model in which the first generation
corresponds to dimension 3, the second to dimension 2, and the third
generation is elementary (dimension 1); and a speculative model
where the effective composite operator dimension is $\sfrac{3}{2}$.
We want to emphasize the fact that the phenomenology is very rich, and is
largely independent of the details of specific models.
More thorough discussions  can be found in Refs. \cite{ALT,SSSB}.

\subsection{SUSY Breaking and Compositeness}
\label{SBC}
We first explain the mechanism that gives rise to SUSY breaking and
compositeness.
The known models have a strong gauge group of the form
$G_{\rm comp} \times G_{\rm lift}$, where both groups are asymptotically
free
and $\Lambda_{\rm comp} \gg \Lambda_{\rm lift}$.
The scale $\Lambda_{\rm comp}$ is the compositeness scale: composite
quarks and leptons become
strongly interacting at the scale $\Lambda_{\rm comp}$.
Direct bounds on the compositeness scale imply that
$\Lambda_{\rm comp} \gsim 2$ TeV.
The role of the gauge group $G_{\rm lift}$ is to generate a dynamical
superpotential that lifts the vacuum degeneracy and gives rise to a local
SUSY breaking minimum.

The models contain the following fields
\beq\nonumber
\begin{tabular}{c|cc|cc}
& $G_{\rm lift}$ & $G_{\rm comp}$  & $G_{\rm global}$ \\
\hline
$Q$ & \fund & \fund & {\bf 1} \\
$L$ &  $\afund$  & {\bf 1}  & $\overline{\fund}$ \\
$\bar{U}$  & {\bf 1} & $\overline{\fund}$ & \fund \\
$P$  & {\bf 1} & $R$ & {\bf 1} \\
\end{tabular}
\eeq
where the representation $R$ may be highly reducible (implying additional
global symmetries).
In addition, the model has a tree-level superpotential
\beq\eql{lambdadef}
W = \lambda Q L \bar{U}.
\eeq
$G_{\rm global}$ must be large enough that it contains  $SU(3)_c \times SU(2)_L
\times U(1)_Y$ as a subgroup.
There are additional requirements on the model in order for this model
to have a local SUSY breaking minimum.
We choose $G_{\rm global}$ such that classically there is a flat
direction with $\bar{U} \ne 0$ where $Q$ and
$L$ are massive and $G_{\rm comp}$ is completely broken.
Nonperturbative $G_{\rm lift}$ dynamics (gaugino condensation)
lift this flat direction via a dynamical superpotential of the form
\beq\eql{wdyn}
W_{\rm dyn} \sim \Lambda_{\rm lift}^{3 - r} \bar{U}^r.
\eeq
Whether this superpotential forces $\bar{U}$ to large or small values depends
on the value of $r$, but it also depends on the effective \Kahler potential
for $\bar{U}$. For $\bar{U} \gg \Lambda_{\rm comp}$,  $G_{\rm comp}$ is broken
at weak coupling and the \Kahler potential is smooth in
$\bar{U}$, so the potential slopes toward the origin for $r > 1$.
For $\bar{U} \ll \Lambda_{\rm comp}$ the $G_{\rm comp}$ dynamics
changes the \Kahler potential for $\bar{U}$.
For example, if the $G_{\rm comp}$ dynamics is confining, the \Kahler
potential will be smooth in terms of a `composite' field $B=(\bar{U}^n)$.
The superpotential can then be written
\beq
W_{\rm dyn} \sim  B^{r/n},
\eeq
which corresponds to a potential that
slopes {\em away} from the origin if $r/n < 1$.
Therefore, for $1 < r < n$ there is no SUSY minimum for any value of
$\bar{U}$, and there is a SUSY breaking minimum near the border between
the region of validity of the confined and Higgs descriptions.
In general, there are other moduli corresponding to excitations of $P$
that must be stabilized;
in some models this requires an additional renormalizable term in the
superpotential, while in other models it requires a dynamical assumption.
The minimum occurs for
\beq\eql{uvev}
\avg{\bar{U}} \sim \frac{\sqrt{N} \Lambda_{\rm comp}}{4\pi},
\eeq
where $N$ is the number of `colors' of $G_{\rm comp}$.

This mechanism also occurs in cases where the $G_{\rm comp}$ dynamics
gives rise to a conformal fixed point (in the limit where we turn off
$G_{\rm lift}$), provided that the $\bar{U}$ anomalous dimensions are
sufficiently large.
As long as $\bar{U} \ll \Lambda_{\rm comp}$ the $G_{\rm comp}$ dynamics is
controlled by the infrared fixed point.
Recall that we are assuming that $G_{\rm lift}$ is weak at the scale
$\Lambda_{\rm comp}$, so the non-perturbative superpotential can be viewed
as a perturbation.
The 1PI potential for $\bar{U}$ is therefore
\beq\eql{conformV}
V_{\rm 1PI} \simeq (K_{\rm 1PI}^{-1})_{\bar{U}^\dagger\bar{U}} \left|
\frac{\d W_{\rm dyn}}{\d\bar{U}} \right|^2,
\eeq
where $K_{\rm 1PI}$ is the 1PI \Kahler metric evaluated at the conformal
fixed point.
The scaling dimension of the \Kahler metric
$(K_{\rm 1PI})_{\bar{U}^\dagger\bar{U}}$
is $2 - 2 d_{\bar{U}}$, where
$d_{\bar{U}}$ is the scaling dimension of $\bar{U}$.
Therefore,
\beq
(K_{\rm 1PI}^{-1})_{\bar{U}^\dagger\bar{U}}
\sim \bar{U}^{2 (d_{\bar{U}} - 1) / d_{\bar{U}}}.
\eeq
This forces the potential to slope away from the origin for
\beq
\frac{1 - d_{\bar{U}}}{d_{\bar{U}}} > r - 1.
\eeq
One might worry that this argument relies on a `Higgs'
description in terms of the elementary field $\bar{U}$ in a regime where
the theory is strongly coupled.
In many cases there is an alternate description in terms of a weakly-coupled
dual theory.
For example, if $G_{\rm comp} = SU(N)$ with $F$ `flavors',
the theory has an infrared fixed point for $\sfrac{3}{2} N < F < 3N$
\cite{SeibergDual}.
There is a dual description in terms of a
theory with gauge group $SU(F - N)$ in which the `baryon'
operator $\bar{U}^N$ in the original
theory is mapped to an operator $\bar{u}^{F - N}$ in the dual.
For $F$ near $\sfrac{3}{2} N$ the dual description is weakly coupled,
and the considerations of the previous paragraph can be made
rigorous.
One finds that the behavior of the \Kahler potential agrees precisely
with \Eq{conformV}.
This equivalence between the `Higgs' and `dual' descriptions can be
viewed as a generalization of the usual `complementarity'   \cite{complement}
for theories with scalars in the fundamental representation, and gives
us additional confidence in the considerations above.

In these models, SUSY is broken by
\beq
\avg{F_{\bar{U}}} \sim \Avg{\frac{\d W_{\rm dyn}}{\d\bar{U}}}
\sim \frac{\Lambda_{\rm comp}^2}{4\pi} ( \lambda \sqrt{N} )^{r - 1}
\left( \frac{\Lambda_{\rm lift}}{\Lambda_{\rm comp}} \right)^{3 - r}.
\eeq
Since $r < 3$ (otherwise the dynamical superpotential \Eq{wdyn}
does not have a good limit
$\Lambda_{\rm lift} \to 0$ when $G_{\rm lift}$ is asymptotically free),
we have $\avg{F_{\bar{U}}} \ll \Lambda_{\rm comp}^2$.
The scalar components of $\bar{U}$ get a SUSY-breaking mass of order
\beq
m^2_{\bar{U}} \sim \Avg{\frac{\d^2}{\d\bar{U}^2} \left|
\frac{\d W_{\rm dyn}}{\d\bar{U}} \right|^2}
\sim \frac{F_{\bar{U}}^2}{\avg{\bar{U}}^2} \equiv m^2_{\rm comp}.
\eeq

The `preon' fields $P$ charged under
$G_{\rm comp}$ get SUSY-breaking masses of order
$m_{\rm comp}$ from effects such as
\beq
\Gamma_{\rm 1PI} \sim \myint d^2\theta d^2\bar{\theta}\,
\frac{16\pi^2}{\Lambda_{\rm comp}^2} \bar{U}^\dagger \bar{U} P^\dagger P
\sim m_{\rm comp}^2 P^\dagger P.
\eeq
The scalar mass squared terms of the preons are not calculable,
and therefore can have either sign.
To build realistic models, we will have to assume that some of
these mass-squared terms are positive.
In the models we discuss, some of the
fermion components of $P$ can remain massless
(in the absence of a Higgs VEV)
because of unbroken chiral
symmetries, and these will be identified with
quarks and leptons.

With this discussion of the mass scales, we have
enough information to analyze the main features of the phenomenology of these
models.
Masses for SM gauginos and elementary charged scalars are
generated by gauge mediation from the composite scalars, so that
\beq
m_{\lambda,{\rm SM}} \sim N \frac{g_{\rm SM}^2}{16\pi^2} m_{\rm comp},
\qquad
m^2_{\phi,{\rm elem}} \sim N \left(
\frac{g_{\rm SM}^2}{16\pi^2} m_{\rm comp} \right)^2.
\eeq
Note that the multiplicity factor $N$ enhances gaugino masses compared to
elementary scalar masses.

In the models we have constructed, some or all of the quarks and leptons from
the first two generations are composite, while the third generation is
elementary.
The reason for this is that in our models
the Yukawa couplings for
composite quarks and leptons arise from higher-dimension operators in the
fundamental theory, and are naturally small compared to one.
It is difficult to accommodate the order-one top Yukawa coupling in this
framework unless the top quark is elementary.
Another reason for the third generation to be elementary is that stop
masses of order $m_{\rm comp} \sim 1$--$10$ TeV (needed to get sufficiently
large gaugino masses) necessitate a large amount
of fine-tuning in electroweak symmetry breaking.
In order to obtain a third generation scalar mass
$m_{3} \gsim 100$ GeV we therefore require
\beq
m_{\rm comp} \gsim \frac{10 {\rm TeV}}{\sqrt{N}}.
\eeq

We see that
single sector models naturally have a superpartner spectrum similar to the
`more minimal' framework \cite{CoKaNe}.
In models of this kind, there is a dangerous negative contribution to the
third-generation squark masses from the heavy scalars \cite{NimaHitoshi}, given
by
\beq\eql{mrg}
\mu \frac{d m_3^2}{d\mu} = \frac{8 g^2}{16\pi^2} C_2
\left[ \frac{3 g^2}{16\pi^2} m_{\rm comp}^2 - m_\lambda^2 \right],
\eeq
where we have assumed that a single gauge group dominates and specialized
to the case of two full composite generations.
One way to avoid this problem is to have the compositeness scale close
to $10$ TeV, so that the negative contribution above does not dominate.
If the compositeness scale is high, one can avoid problems if the gaugino
contribution is important. From \Eq{mrg}, we see that
$m_{\lambda} \gsim m_{\rm comp} / 10$ is sufficient.
This requires $N \gsim 10$.

Most of the models we have constructed have of order $N$ `preonic'
generations above the
compositeness scale, and for $N \gsim 10$ the SM gauge groups
are far from asymptotically free.
This is compatible with perturbative unification if the compositeness scale
is above (or near) the GUT scale $10^{16}$ GeV.

\subsection{`Meson' Models}
We will now  discuss `meson' models where all quarks and leptons of the first
two
generations correspond to dimension-2 operators $ P \bar{U}$
in the fundamental theory.
This means that small Yukawa couplings involving the first two generations can
be
generated by:
\beq\eql{massop}
W_{\rm Yuk} =
\frac{1}{M^2} H (P \bar{U}) (P \bar{U})
+ \frac{1}{M} H \Phi_3 (P \bar{U}).
\eeq
where $M$ is the scale of new physics where flavor symmetries are broken,
$H$ is a Higgs field, and $\Phi_3$ is an elementary third-generation
quark or lepton field.
This gives a Yukawa matrix of the form
\beq
y \sim \pmatrix{
\epsilon^2 & \epsilon^2 & \epsilon \cr
\epsilon^2 & \epsilon^2 & \epsilon \cr
\epsilon   & \epsilon   & 1   \cr},
\qquad
\epsilon \sim \frac{\avg{\bar{U}}}{M},
\eeq
Additional structure is clearly needed to construct fully realistic
Yukawa matrices, but for $\epsilon$ in the range $10^{-1}$--$10^{-2}$
this is a good starting point.

We will make the conservative assumption that the new physics at the scale
$M$ does not have any approximate flavor symmetries that can suppress FCNC's.
In particular, this means that the Yukawa couplings $\lambda$ in \Eq{lambdadef}
do not conserve flavor.
It is highly non-trivial that the strong dynamics in this theory nevertheless
gives rise to an approximate flavor symmetry at low energies that enforces
the near degeneracy of the composite scalars.
The underlying reason for this is the fact that all of the composites
$(P \bar{U})$ are part of a single multiplet from the point of view of the
strong interactions.

Let us first consider the $\lambda$-dependent effects.
The superpotential \Eq{wdyn} depends on $\lambda$ only through
$\det(\lambda)$, which is flavor independent.
There is nontrivial $\lambda$ dependence in the effective \Kahler potential,
but it is proportional to $\lambda^2 / (16\pi^2) \lsim 10^{-2}$.
We now consider the effects of general higher-dimension operators
suppressed by the flavor scale $M$.
The largest effects come from terms in the effective Lagrangian
of the form
\beq\eql{MFCNC}
\Delta\scr{L}_{\rm eff} \sim \myint d^2\theta d^2\bar{\theta}\,
\frac{1}{M^2} (P \bar{U})^\dagger (P \bar{U}),
\eeq
which give rise to mixing between the composite generations.
This translates to mixing masses between the composite generations of order
\beq
\frac{\Delta m^2_{jk}}{m_{\rm comp}^2} \sim \left(
\frac{\avg{\bar{U}}}{M} \right)^2 \sim y_{jk}.
\eeq
The most stringent bounds on squark mixing come from $K^0$--$\bar{K}^0$
mixing, and can be summarized as
\beq\eql{KKbar}
\Re \left( \frac{\Delta m^2_{\tilde{d}\tilde{s}}}{m_{\rm comp}^2} \right)
\lsim 10^{-1} \frac{m_{\rm comp}}{10 {\rm TeV}},
\qquad
\Im \left( \frac{\Delta m^2_{\tilde{d}\tilde{s}}}{m_{\rm comp}^2} \right)
\lsim 10^{-2} \frac{m_{\rm comp}}{10 {\rm TeV}}.
\eeq
Since $y_{ds} \sim 3 \times 10^{-4}$, this is easily satisfied even if we
assume that $C\!{}P$ violation in the flavor sector is maximal.

A striking signature of these models is that {\em all} scalars of the
first two generations unify at the scale $\Lambda_{\rm comp}$ (which need
not be close to the GUT scale).
The unification holds up to effects suppressed by a loop factor, and
is therefore expected to hold to $1\%$.
This striking pattern is difficult to obtain naturally in other SUSY breaking
models.

An explicit model of this type was constructed in \cite{SSSB}
by identifying $G_{\rm lift} \times G_{\rm comp} \times G_{\rm global}$
with $SU(13)_{\rm lift} \times SU(15)_{\rm comp} \times SU(15)_{\rm gl}$ and
taking
the preon field $P$ to consist of  3 $\fund$'s and  one $\overline{\fund}$
of $SU(15)_{\rm comp} $, where one $\fund$ of $SU(15)_{\rm gl}$ decomposes
into one SM generation.

\subsection{`Dimensional Hierarchy' Models}
We next discuss `dimensional hierarchy' models that explain the observed
fermion mass hierarchy in terms of a hierarchy of dimensions of operators.
Specifically, we assume that the first-generation quarks and leptons correspond
to dimension 3 operators of the form $(P \bar{U} \bar{U})$,
second-generation quarks and leptons correspond to dimension 2 operators
$(P \bar{U})$, and third generation quarks and leptons are elementary
(dimension 1).
In this case, Yukawa couplings involving the composite states arise from
terms in the tree-level superpotential of the form
\beq
W_{\rm Yuk} &= \frac{1}{M^4} H (P \bar{U} \bar{U}) (P \bar{U} \bar{U})
+ \frac{1}{M^3} H (P \bar{U} \bar{U}) (P \bar{U})
+ \frac{1}{M^2} H \Phi_3 (P \bar{U} \bar{U})
\\
&\qquad\qquad
+ \frac{1}{M^2} H (P \bar{U}) (P \bar{U})
+ \frac{1}{M} H \Phi_3 (P \bar{U}),
\eeq
giving rise to a Yukawa matrix of the form
\beq
y \sim \pmatrix{
\epsilon^4 & \epsilon^3 & \epsilon^2 \cr
\epsilon^3 & \epsilon^2 & \epsilon   \cr
\epsilon^2 & \epsilon   & 1     \cr},
\qquad
\epsilon \sim \frac{\avg{\bar{U}}}{M}.
\eeq
This structure reproduces the main features of the observed fermion
mass hierarchy for $\epsilon \sim 10^{-1}$.
A striking signature of these models is that the first- and second-generation
scalars unify in two multiplets at the scale $\Lambda_{\rm comp}$.

In this scenario there is no approximate flavor symmetry at low
energies because the first- and second-generation fields belong to different
strong-interaction multiplets.
We therefore have
\beq
\frac{\Delta m^2_{\tilde{d}\tilde{s}}}{m_{\rm comp}^2} \sim \sin\theta_{\rm c}
\sim 10^{-1}.
\eeq
Comparing with the bounds from the $K^0$--$\bar{K}^0$ system \Eq{KKbar},
we see that $m_{\rm comp} \sim 10$ TeV is sufficient to suppress FCNC's but
we require either a $10\%$
fine-tuning or a $10\%$ suppression of $C\!{}P$-violating effects in the squark
masses.

An explicit model of this type was constructed in \cite{SSSB}
by identifying $G_{\rm lift} \times G_{\rm comp} \times G_{\rm global}$
with $SU(2)_{\rm lift} \times SU(16)_{\rm comp} \times SU(16)$ and taking
the preon field $P$ to consist of  2 $ \fund$'s and  one antisymmetric
tensor of $SU(16)_{\rm comp} $.

\subsection{ A Simple but Speculative Model}
We now present a simple model
whose dynamics we do not know how to analyze completely.
If we make a reasonable dynamical assumption, this model gives rise to
compositeness and SUSY breaking by the mechanism discussed in above.
The particle content is:
\beq\nonumber
\begin{tabular}{c|cc|cc}
 & $SU(k)$ & $SO(10)$ & $SU(10)$ & $SU(2)$  \\
\hline
$Q$ & $\fund$ & $\fund$ & {\bf 1} & {\bf 1}   \\
$L$ & $\afund$ & {\bf 1} & $\fund$ & {\bf 1}   \\
$\bar{U}$ & {\bf 1} & $\fund$ & $\afund$ & {\bf 1}   \\
$P$ & {\bf 1} & ${\bf 16}$ & {\bf 1} & $\fund$   \\
\end{tabular}~.
\eeq

For $\avg{\bar{U}} \gg \Lambda_{10}$,
$SO(10) \times SU(10)$ is broken to the diagonal $SO(10)$ subgroup and
$SU(k)$ gaugino condensation gives rise to a dynamical superpotential
\beq\eql{SOwdyn}
W_{\rm dyn} \sim \bar{U}^{10/k}.
\eeq
The potential therefore slopes toward $\bar{U} \to 0$ for $k < 10$.

The dynamics for small values of $\avg{\bar{U}}$ involves the strong-coupling
behavior of the $SO(10)$ gauge theory with spinors, which is presently
not well understood.
The $SO(10)$ gauge theory has a dual description in terms of an
$SU(2) \times SU(7+k)$ gauge theory  \cite{SO10duals};
this dual is not weakly coupled in the infrared, so we cannot use it
to determine the behavior of the \Kahler potential for $\bar{U}$.
Based on analogies with similar duals, one expects this theory to have
an infrared fixed point \cite{SO10duals,fiveeasy}.

If we assume that the anomalous dimension of $\bar{U}$ is sufficiently
large ($> 0.1$),
then there is a local SUSY-breaking minimum.
The fermions from the {\bf 16}'s are exactly massless far from the origin, and
because there can be no phase transitions as a function of moduli they
are massless at the local minimum as well.
This model therefore contains two composite fermionic {\bf 16}'s which can
be identified with two SM generations (with right-handed neutrinos)
if we embed the SM into $SU(10)$ via the standard $SO(10)$ GUT
embedding
FCNC's are suppressed by the approximate global $SU(2)$ symmetry of the
strong dynamics.
Above the compositeness scale, this model has
$3 + k/2$ additional `preonic' generations.

Yukawa couplings for the composite generations can be induced if we include
a Higgs field, $H$ embedded in the $\fund$ of the global $SU(10)$ by operators
of the form
\beq
W_{\rm Yuk} =
\frac{1}{M} P P H  \bar{U}
\eeq
This gives Yukawa couplings $y \sim \avg{\bar{U}} / M$ for the composite
quarks and leptons.
(Compared with our previous expressions,
this corresponds to the composite operators
being dimension $\sfrac{3}{2}$.)
Thus the flavor scale $M$ can be pushed up even higher in
this model, and FCNC's are even more suppressed than in our
`meson' models.

\section{Conclusions}
We have seen that there is a wide class of realistic
models which dynamically break SUSY
and produce composite quarks and leptons,
all in a single (strongly-coupled) sector.
These models are remarkably simple;
many of the fermion mass hierarchies follow naturally, and approximate flavor
symmetries of the strong interactions can guarantee the natural suppression
of flavor-changing neutral currents (including $\epsilon_K$) with no
fine-tuning.

All of these models have a very distinctive phenomenology:
the composite sfermions of the first two generation are heavier than the
gauginos, Higgsinos, and third-generation sfermions;
and the composite sfermions unify at the compositeness scale.

A (perhaps) surprising result of our recent work \cite{SSSB}
is the wide variety of dynamics that can give
rise to simultaneous compositeness and SUSY breaking.
We have found models whose low-energy dynamics is governed by
either
confinement, a non-trivial infrared fixed point, or a free-magnetic
phase.
We believe that it is quite likely that further progress in understanding
the dynamics of SUSY gauge theories will lead to the discovery of many
additional models that display the dynamics illustrated here.

\section{Acknowledgments}
We thank N. Arkani-Hamed,
C. Cs\'aki, R. Rattazzi, H. Murayama, A. Nelson, and M. Schmaltz
for discussions.
M.A.L. is supported by a fellowship from the Alfred P. Sloan Foundation.
J.T. is supported by the National Science
Foundation under grant PHY-95-14797, and is also partially supported by
the Department of Energy under contract DE-AC03-76SF00098.

\end{document}